%% file: rsp4cha.tex
\documentclass[aps,twocolumn,superscriptaddress,floatfix,amsmath,a4paper]{revtex4}
\usepackage{graphicx}
\usepackage{amsmath,amssymb}
\usepackage[utf8]{inputenc}
\usepackage{psfrag}
\usepackage{bm}
\newcommand{\chic}{\chi_{\scriptstyle \rm c}}
\newcommand{\Qc}{Q_{\scriptstyle \rm c}}

\begin{document}
\title{Globally synchronized oscillations in complex cyclic games}

\author{Charlotte Rulquin}
\email{rulquin@clipper.ens.fr}
\affiliation{École Normale Supérieure,  International Center of Fundamental Physics, 45 Rue d'Ulm, 75005 Paris, France}
\affiliation{Instituto de Física, Universidade Federal do Rio Grande do Sul, CP 15051, 91501-970 Porto Alegre RS, Brazil}
\author{Jeferson J. Arenzon}
\email{arenzon@if.ufrgs.br}
\affiliation{Instituto de Física, Universidade Federal do Rio Grande do Sul, CP 15051, 91501-970 Porto Alegre RS, Brazil}

\date{\today}

\begin{abstract}
The Rock-Paper-Scissors (RPS) game and its generalizations with ${\cal S}>3$ species
are well studied models for cyclically interacting populations. 
Four is, however, the minimum number of species that, by allowing other interactions 
beyond the single, cyclic loop, breaks both the full intransitivity of 
the food graph and the one predator, one prey symmetry. Lütz {\it et al}
(J. Theor. Biol. {\bf 317} (2013) 286) have shown the existence, on a square
lattice, of two  distinct phases, with either four or three coexisting species.
In both phases, each agent is eventually replaced by one of its predators but
these strategy oscillations remain localized as long as the interactions are short ranged.
Distant regions may be either out of phase or cycling through different food web subloops 
(if any). Here we show that upon replacing a minimum fraction $Q$ of the short range 
interactions by long range ones, there is a 
Hopf bifurcation and global oscillations become stable. Surprisingly, to build such 
long distance, global synchronization, the four species coexistence phase requires less 
long range interactions than the three species phase, while one would naively
expect the contrary. Moreover, deviations from highly homogeneous conditions ($\chi=0$
or 1) increase $\Qc$ and the more heterogeneous is the food web, the harder the
synchronization is. By further increasing $Q$, while the three species phase remains 
stable, the four species one has a transition to an absorbing, single species state.  
The existence of a phase with global oscillations for ${\cal S}>3$,
when the interaction graph has multiple subloops and several possible local 
cycles, lead to the conjecture that global oscillations are a general characteristic,
even for large, realistic food webs.
\end{abstract}

\maketitle

\section{Introduction}

Actual food webs are usually large and complex graphs and although their structure 
alone is not always enough to predict whether extinctions will occur or which 
species will eventually dominate, understanding the role of the specific local 
modules (hierarchical, tree-like, loop, dangling ends, etc) is important 
when modeling competing populations. 
The existence of closed, oriented loops, in which the ${\cal S}$ species are not ranked, 
is believed  to be an important mechanism fostering persistent coexistence in 
the presence of competition among species. In this case, each species is eventually 
replaced by a predator in a cyclic manner~\cite{HoSi98,SzFa07}. 
For ${\cal S}>3$, several partially overlapping cycles 
are possible, and the larger ${\cal S}$ is, the harder it is to identify the whole 
ensemble of interactions. Indeed, observing these trophic networks in 
real systems, and resolving the effects of its topology and internal dynamics 
from noisy external causes is not a simple task. It is thus not surprising that most of the observed 
examples are for small ${\cal S}$~\cite{Gilpin75,Tainaka88}: mating lizards~\cite{SiLi96}, competing 
bacteria~\cite{KeRiFeBo02,KiRi04,HiFuPaPe10,Trosvik10}, coral reef environments~\cite{BuJa79}, 
competing grasses~\cite{Watt47,Thorhallsdottir90,SiLiDa94}, etc. It remains necessary, tough,
to systematically study which properties are robust when ${\cal S}$ is large and which are
specific of systems with a small number of species.

For ${\cal S}=3$, the simplest and well studied 
Rock-Scissors-Paper (RSP) game, the interaction graph is a three vertices, single loop
oriented graph in which all three species interact with each other.
A direct generalization~\cite{FrKrBe96,FrKr98,SaYoKo02,SzSz04b,SzSzSz07,CaDuPlZi10,DuCaPlZi11,RoKoPl12}
considers an oriented ring with ${\cal S}>3$, $0\to 1\to \ldots \to {\cal S}-1\to 0$
and, while nearest neighbors species along the ring interact, the others
are mutually neutral. In particular, ${\cal S}=4$ is the simplest case in which
neutral pairs may form non interacting, passive alliances that prevent or 
delay invasions~\cite{SaYoKo02,SzSz04b,Szabo05,SzSzSz07,SzSz08,CaDuPlZi10,DoFr12,DuCaPlZi11,RoKoPl12,GuLoGi13,InPl13}.
On the other hand, active defensive alliances~\footnote{Differently from a passive defensive
alliance, the species belonging to an active alliance also predate on each other, but
prevent species not belonging to the subcycle from invading.} 
may appear among
non mutually neutral species (cyclic alliances) when the interaction graph has more than a single 
loop~\cite{SiHoJoDa92,DuLe98,SzCz01a,SzCz01b,Szabo05,KiLiUmLe05,PeSzSz07,SzSzBo08,LaSc08,LaSc09,HaPaKi09,LiDoYa12,AvBaLoMe12,AvBaLoMeOl12,RoDaPl13,LuRiAr13}.
This is obtained, for ${\cal S}=4$, by introducing crossed
interactions with rate $\chi$ (see Fig.~\ref{fig.graph}),
that turn the ring into a fully connected graph and break not only its full intransitivity~\cite{LuRiAr13,KnKrWeFr13} 
but also the equal number of predators and preys symmetry. In spite of that, this model still
presents a four species coexistence phase
for not too large values of $\chi$: beyond a threshold $\chic\simeq 0.35$, the weakest species 
gets extinct and the food web changes to an inhomogeneous RSP game. 

While the system had not yet attained an absorbing state, local oscillations are observed 
as each agent is eventually replaced by one of its predators. That is, each site is
a simple oscillator as strategies cycle in a sequence obbeying the food web.
In the square lattice, two of these oscillators, located at sites far apart, are not able to 
have these oscillations in phase. Indeed, the short range interactions are not able to build correlations involving
large portions of the system and when the global densities are evaluated, these phase
differences average out and the system attains a fixed point (up to finite size fluctuations), that is, 
no oscillations in global densities are observed. In analogy to epidemiological models~\cite{KuAb01},
upon the introduction of long range interactions for ${\cal S}=3$, since the average path
length between sites becomes much smaller, global synchronization 
becomes possible and there is a Hopf bifurcation from the fixed point to a
limit cycle~\cite{SzSzIz04,KiLiUmLe05,YiHuWa07}. For more complex food webs, ${\cal S}>3$, with the simultaneous presence of 
several possible loops, and the possibility that distinct regions
of a spatially distributed system follow different subcycles, it is not a priori obvious whether well 
separated regions would be able to synchronize. This poses the question of the very existence of the Hopf
bifurcation and whether global oscillations may be stable or not under  the presence of several
possible local cycles. Would they require a stronger correlation between distant regions and, as
a consequence, a larger amount of long range connections? 
%Indeed, once unconstrained long range interactions are introduced, the formation of
%spatial domains of allies or invading spiral waves are affected as it becomes
%possible to directly access some agents inside these regions. 
Moreover, whether there is a larger or smaller species diversity when long range
interactions are introduced seems to strongly depend on the details
of the updating rule chosen to model the individual interactions~\cite{LaSc09,RoAl11}. Thus,
further important points are to understand to what extent the coexistence state is robust against
both the change in the range of the interactions and the presence of global oscillations 
and how these results may be extended to other ${\cal S}>3$ systems.

 In order to address the above questions, we study the ${\cal S}=4$ system introduced
in Ref.~\cite{LuRiAr13} and consider a simple network that
interpolates between the local, square lattice, and the non local random
graph. In Sec.~\ref{section.model} we introduce the model and describe the topology
of the food web and the spatial network. Then, in Sec.~\ref{section.results}
we present our results and, finally, discuss and present our conclusions in
Sec.~\ref{section.conclusions}.

\section{Model, dynamics and network topology}
\label{section.model}

\begin{figure}[htp] 
\includegraphics[width=5cm]{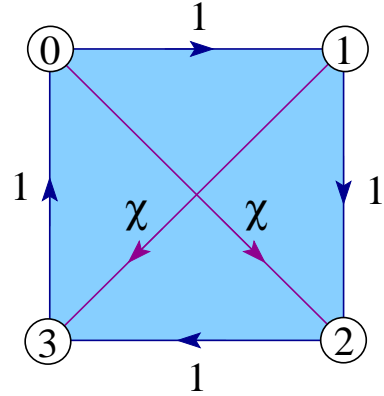}
\caption{Species interactions represented by an oriented graph, arrows indicating the 
invasion direction and 1 and $0\leq \chi\leq 1$ are the corresponding rates~\cite{LuRiAr13}.}
\label{fig.graph}
\end{figure}

The starting model consists of a fully intransitive system of four species (or strategies),
identified from 0 to 3, such that the species $i$ outcompetes the species $(i+1)\, {\tt mod}\, 4$,
that is, $0 \rightarrow 1\rightarrow 2\rightarrow 3\rightarrow 0$ (0123, for short)
with uniform, unitary rate. We break the mutual neutrality of species $(0,2)$ and $(1,3)$ by
adding crossed interactions with a tunable rate $0\leq\chi\leq 1$ (Fig.~\ref{fig.graph}).
Whenever $\chi\neq 0$, some hierarchy is present since the subloops 123 and 012 are transitive while
013 and 023 remain intransitive. Also, the symmetry of the graph is broken: while all species are equivalent if $\chi=0$ (one prey,
one predator each),  for $\chi \neq 0$ they separate into two groups: species 0 and 1 have two preys and one predator, while species 2 and 3 have two predators and one prey. This strong hierarchy is attenuated by the fact that, even if species 2 and 3 have only one prey each, they always subjugate them, while species 0 and 1 can sometimes fail if $\chi<1$.
Indeed, several recent works (e.g., Refs.~\cite{SzSz08,LuRiAr13,VuSzSz13,KnKrWeFr13}) 
have shown that the structure of the interaction graph alone is not 
enough to predict the asymptotic temporal evolution of the system, 
stressing the importance of the invasion rates. 

Since we are mainly concerned with the effects of spatial correlations, the model will be studied,
through Monte Carlo simulations, on a regular graph with both short and long range interactions and
a constant number $N$ of individuals.
We start with a square lattice and periodic boundary conditions and, without changing the coordination
number, replace a fraction $Q$ of randomly chosen short range connections with long range ones
following the rewiring process detailed in Ref.~\cite{SzSzIz04}. 
While intermediate $Q$ produces a small world network, $Q=0$ recovers the square lattice
and $Q=1$ a random regular graph or Bethe lattice (since typical loops, involving $\log N$ sites, become 
irrelevant in the thermodynamical limit). Starting from the square lattice, we choose a site and 
one of its nearest neighbors. The link between them is opened and
reconnected to another randomly chosen, non nearest-neighbor site (self interactions are excluded). This last site now has an extra connection and, in order to keep the coordination constant, we remove one of its local connections. 
We now take the site with one neighbor missing and reconnect it with another randomly chosen site. The
process is repeated until the fraction $Q$ is reached and the graph is closed by
connecting the last link freed to the missing link site of the first step. 
Once the rewiring finishes, no further modification in the network occurs,
that is, it is quenched. There are, of course, other ways of introducing
long range interactions in the system. We chose, however, this rewiring 
algorithm, that replaces short by long range interactions, in order to
a) reproduce previous results, b) keep the graph regular (assuming that in 
a time step, each agent performs a constant number of interactions) and 
c) interpolate between the square lattice and the random regular graph.

The system is equally populated with all four strategies at time $t=0$.
During the time evolution, one site and one of its four neighbors are randomly chosen and their 
strategies are updated following the graph of Fig.~\ref{fig.graph}: the predator
replaces the prey with the corresponding rate, either 1 or $\chi$. The time unit, one Monte Carlo 
step (MCS), corresponds to $N$ such updating attempts. We measure
the density of each strategy along the trajectory and the results below are obtained from
these temporal series. This purely competitive, serial dynamics does not take 
into account other relevant ingredients like decoupling the reproduction and predation steps, swapping or
displacement of agents, mutation, etc, although it could be extended to included them~\cite{SzFa07}.

\section{Results}
\label{section.results}

Finite systems following the dynamics described in the previous section
present a sequence of extinction events, eventually attaining an absorbing, 
homogeneous state with a single species. These extinctions, by removing
nodes and links, also change the food web, making it simpler. Nonetheless, the dependence 
of a conveniently defined average extinction time $\tau$ on the system size $N$ is
useful to characterize the different dynamical phases of the model~\cite{AnSc06,ReMoFr07a,CrReFr09b,ScCl10b}. 
These phases have diverse levels of diversity with either one, three or four species coexisting~\cite{LuRiAr13}.
For the transition from four to three coexisting species phases, the
timescale of the first extinction is the relevant one~\cite{LuRiAr13}, while for the
transition to an homogeneous state is the last extinction that matters.
The coexistence is said to be stable when the related deterministic dynamics presents
a stable attractor in the coexistence phase, and this is associated with an exponentially
increasing time for the first extinction away from the coexistence state to occur as $N$ increases.  Analogously, the
unstable state presents only a logarithmic increase of the extinction time and the deterministic
system approaches an absorbing state. In between, a power law dependence of the extinction
time on the system size is related with the presence of closed, neutrally stable orbits
in the deterministic case. 

\begin{figure}[htp] 
\includegraphics[width=4cm]{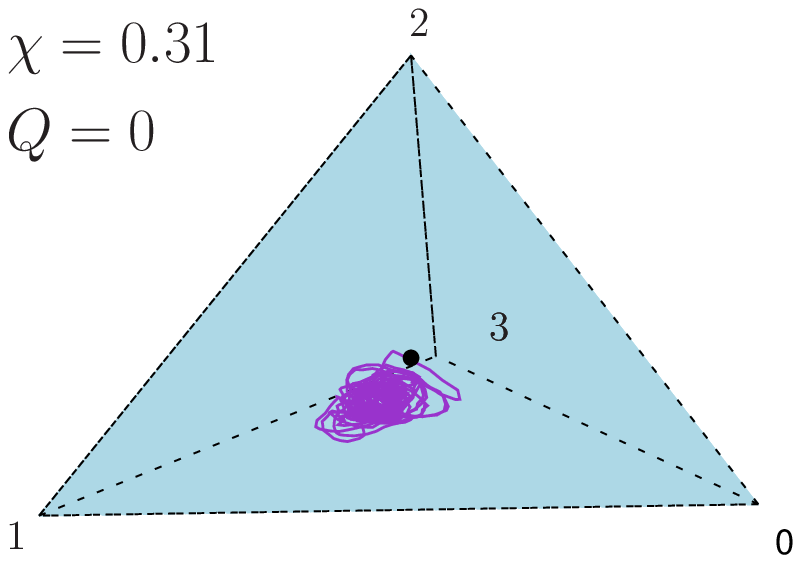}
\includegraphics[width=4cm]{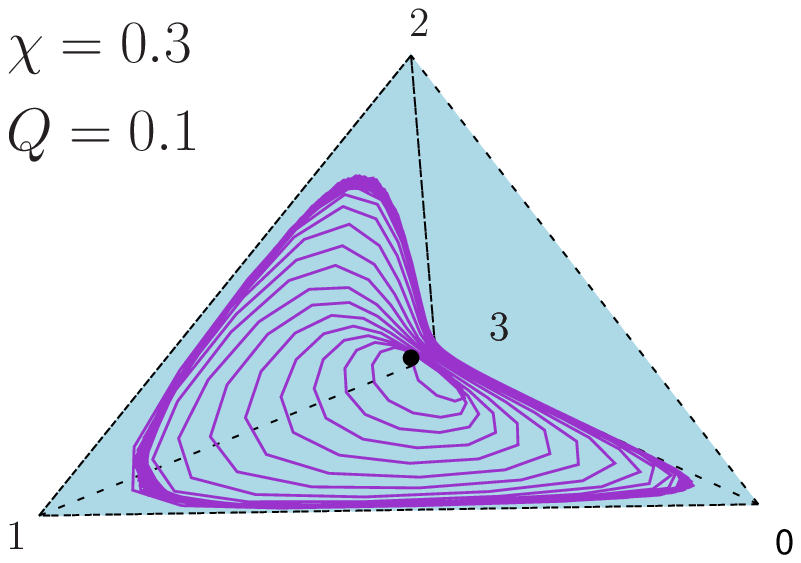}
\begin{center}
(a) \hspace{3.5cm} (b)
\end{center}

\includegraphics[width=4cm]{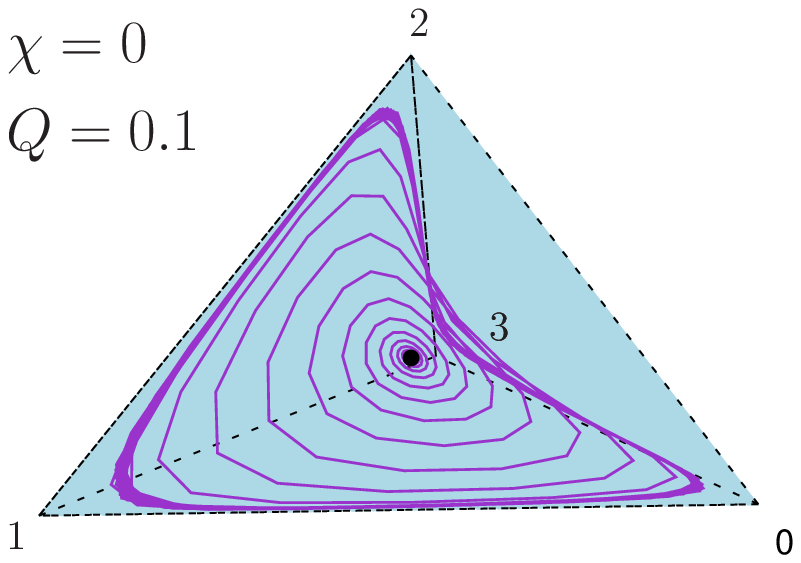}
\begin{center}
(c)
\end{center}

\caption{Simplex representation of the time evolution of the four densities for several values of 
$Q$ and $\chi$, showing either local or global oscillations (0123).
% (otherwise the orbit would be confined to one of the sides of the simplex). 
The initial condition (black dot), is the 
homogeneous state, $\rho_i=1/4, \forall i$ and the system size $N$ is large enough to prevent
strong finite size effects. 
a) $Q=0$ and $\chi = 0.3 \leq \chic$ for $N=10^4$: finite size, stochastic fluctuations
around the asymptotic fixed point. The amplitude of these fluctuations decreases
with $N$~\cite{LuRiAr13} and all oscillations are localized. 
b) $Q=0.1$ and $\chi=0.3 \leq \chic$ for $N=4\times 10^6$: oscillations are now global and the asymptotic
state is a limit cycle whose perimeter may be used as an order parameter for the transition.
c) $Q=0.1$ and $\chi = 0$ for $N=4\times 10^6$: similar to (b) but for a fully intransitive
system: the orbit approaches the heteroclinic one and the average perimeter $\phi$ is close 
to the maximum value. 
%c) $Q=0.3 \geq Q_C^1$ and $\chi=0$ for $N=2.5\times 10^7$: the perimeter $P$ tends to a value very near from 1, and in this limit the fluctuations due to the finite size of the system are very important. The exact behavior of the system above this value of $Q$ remains unknown, technical computer limits prevent us to apply our simulations for bigger sizes, and new analyses will be necessary to characterize this region.
}
\label{fig.simplex}
\end{figure}

\begin{figure}[htp]
\includegraphics[width=8cm]{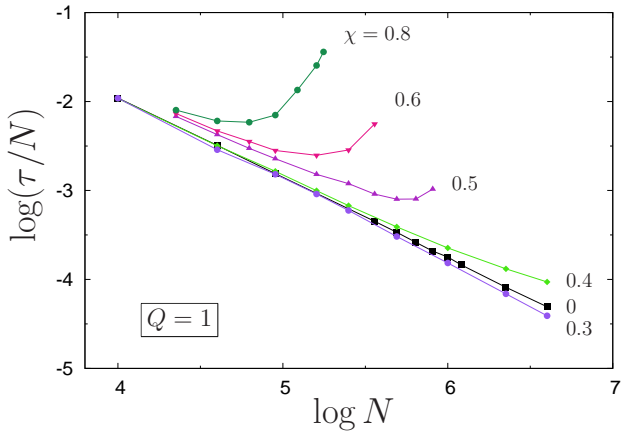}
\caption{Log-log plot of the rescaled average last extinction time, $\tau/N$ {\it versus} $N$ for $Q=1$
(random graph). For this case, the system has two phases, with three or one surviving
species and $\tau$ is the characteristic time to attain an absorbing state.
Notice the presence, for large  
$\chi$, of a minimum of $\tau/N$. Before this minimum, $\tau$ depends 
sub-linearly on $N$ but crosses over to an exponential growth at sufficiently large sizes. 
For smaller values of $\chi$, at least within the sizes and times considered here,
no evidences of such a minimum were observed. At late times, 
there are two different regimes: if $\chi \geq 0.4$, $\tau \propto \exp N$, 
while for $\chi \leq 0.3$, $\tau \propto \log N$. Thus, the critical value seems to lie in the interval 
$0.3 < \chi_c < 0.4$.}
\label{fig.tau_q1_chi}
\end{figure}

Let us first compare the two extreme cases, $Q=0$ (square lattice) and $Q=1$ (random
graph). The former was studied in Ref.~\cite{LuRiAr13} and presents two coexisting
phases without global oscillations, with either four ($\chi<\chic\simeq 0.35$) or 
three species ($\chi>\chic$). In these phases, the timescale of the first extinction 
grows, respectively, exponential or logarithmically with $N$. An example of the time 
evolution, starting from the homogeneous state, $\rho_i=1/4\; \forall i$ (black dot), 
is shown in Fig.~\ref{fig.simplex}a. The orbit is confined to a small region around 
the attracting fixed point (the small volume is due to finite size fluctuations). 
For the random graph ($Q=1$), on the other hand, species 2 always becomes extinct 
in the early stages of the dynamics (both species 2 and 3 have two predators, but
those of species 2 have two preys and are strong) and the possible asymptotic phases have one or 
three species, for small or large $\chi$, respectively. The relevant timescale is the 
time to attain the absorbing state, and it is shown in Fig.~\ref{fig.tau_q1_chi}. 
Above the threshold, %differently from the behavior on the square lattice~\cite{LuRiAr13}, 
after an initial sublinear dependence on $N$ (up to a crossover size, the minima in 
Fig.~\ref{fig.tau_q1_chi}, for $\chi \geq 0.4$), $\tau$ enters an exponential regime.
The location of the minimum grows very fast as $\chi$ approaches the threshold
and seems to diverge for some value in the interval $0.3<\chi<0.4$. For smaller
values of $\chi$, there is no longer a minimum.  This three species phase for 
large $\chi$ also occurs for $Q=0$. Thus, for large $\chi$, it extends
over the whole interval, from $Q=0$ to 1. The level of transitivity in this case is too high
to sustain four species and the food web turns into an heterogeneous 
RSP game (013 with rates 1, $\chi$ and 1, respectively). For small $\chi$, on the
other hand, the behavior is reminiscent of $\chi=0$, in which only one or 4 species 
coexistence are observed since there is no cross interactions (although, in principle,
two mutually neutral species are also possible).
In this phase, the growth of $\tau$ is rather slow (logarithmic) and, because extinctions 
are probable, the asymptotic state is the homogeneous, single species phase. Interestingly, while for $Q=0$ 
the system loses diversity when increasing $\chi$, passing from four to three 
species, for $Q=1$ it is the opposite, from the homogeneous single species to 
three coexisting ones. Diversity decreases in the former and increases in the latter. 

\begin{figure}[htp]
\includegraphics[width=8cm]{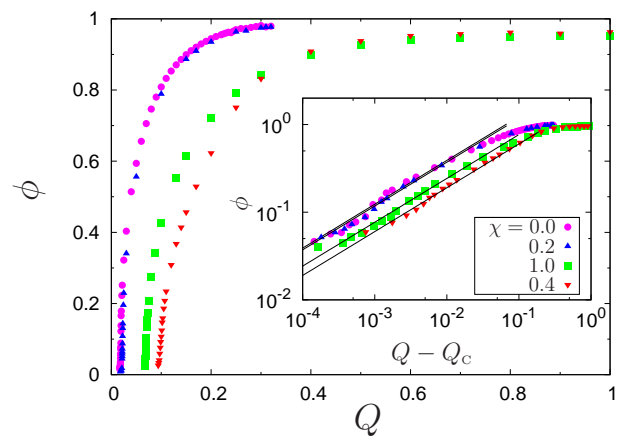}
\caption{Perimeter $\phi$ of the limit cycle as a function of $Q$ for several values
of $\chi$. The point at which
$\phi\to 0$, $Q_{\scriptscriptstyle\rm c}(\chi)$, locates the Hopf bifurcation from
local to global oscillations. Inset: log-log plot close to the transition showing
that the exponent, 1/2, does not depend on $\chi$.}
\label{fig.perimeter}
\end{figure}

Even more interesting is the behavior for a fixed $\chi$ as $Q$ increases.
There is a Hopf transition at $\Qc$ from the fixed point existing 
at small $Q$ to an asymptotically stable limit cycle. Two examples are 
shown in Fig.~\ref{fig.simplex}b and c. In both cases, with both $\chi=0$ 
and 0.3, the orbit approaches the four vertices in the order 0123, following 
the outer loop of the food web. In the intransitive case, $\chi=0$, 
Fig.~\ref{fig.simplex}c, the actual path approaches the heteroclinic orbit
and the system is close to the transition to the absorbing state. These
global oscillations, for not so high $Q$, are observed either both in the 
four and three species phases: while in the former the orbit is in the 
interior of the simplex, in the latter it is confined to one of its faces. 
Once in the asymptotic state, one can measure the perimeter $\phi(Q)$ of each cycle
(between two consecutive crossings of the point representing the state of
the system through a conveniently defined Poincaré
section inside the simplex, we add up its displacement at each MC step). This is averaged over the whole evolution 
of the system and normalized by the perimeter 
of the heteroclinic orbit such that $0\leq \phi < 1$. Fig.~\ref{fig.perimeter}
shows that the amplitude of the oscillations increases continuously with $Q$ 
for different values of $\chi$, analogously to the ${\cal S}=3$ case~\cite{SzSzIz04}.
Above the threshold $\chic$, that is, inside the three species phase, $\phi$ remains
smaller than its maximum value for all values of $Q$ up to the random graph 
limit. In other words, the global oscillation phase extends up to $Q=1$, 
consistently with Ref.~\cite{SzSz04a}. On the other hand, the behavior
of $\phi$ inside the four species global oscillations phase, at small $\chi$, is 
different as $\phi$ gets very close to unity, corresponding to an almost 
heteroclinic orbit (see Fig.~\ref{fig.simplex}c), very close to the faces of the 
tetrahedron. Although it is not easy to precisely locate the transition to
the absorbing state as very large systems are needed in this case, our 
results indicate that it indeed occurs for $Q$ not much larger than 0.3.
% Slightly above this value the system goes through a sequence of 
%extinctions and attains the homogeneous absorbing state. 
Thus, in this case the global phase exists only for intermediate 
values of $Q$ for which $0<\phi<1$. A final remark concerns the exponent with
which $\phi$ goes to 0 at the transition from local to global oscillations.
In the inset of Fig.~\ref{fig.perimeter}, for several values of $\chi$,
the declivity seems to be the same close to the transition and the exponent is 1/2,
that is, $\phi\sim (Q-\Qc)^{1/2}$, consistent with a Hopf bifurcation.
The universality of the exponent is important: not only it does not seem to 
depend on the original lattice structure upon which long range connections were 
added but also on the complexity of the global cycle after the transition. Whether this universal
behavior is valid for even larger cycles, when ${\cal S}>4$, is an open question.

\begin{figure}[htp]
\includegraphics[width=8cm]{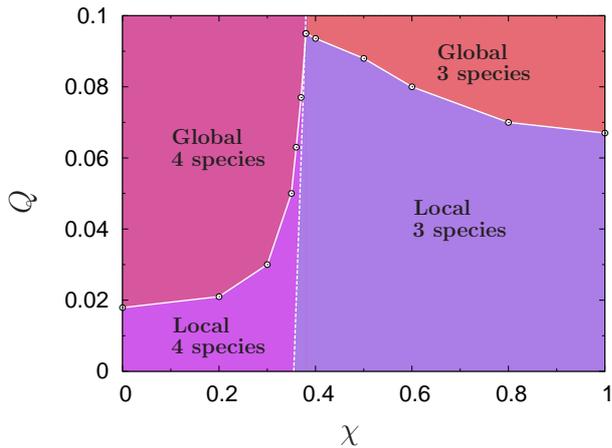}
\caption{Phase diagram. The behavior at small values of $Q$ follows
the $Q=0$ case (square lattice) studied in Ref.~\cite{LuRiAr13} and has a
transition from a four  to three species coexistence state. Since
there is only short-range interactions, only local oscillations are present in both
phases. 
There is a minimum amount of long-range connections,  $\Qc(\chi)$, necessary to 
synchronize these local oscillations throughout the system. Interestingly, while the 
system presents coexistence of all four species (and subloops in the food web are relevant),
$\Qc$ is smaller than in the three species phase. For even larger values of $Q$,
the four species phase is replaced by a less interesting single species, homogeneous phase.}
\label{fig.diagram}
\end{figure}

The phase diagram in Fig.~\ref{fig.diagram} summarizes the above results.
When the amount of long range connections is small (low $Q$) only local
oscillations are possible, whatever is the number of species in the
coexistence state. Upon increasing $Q$ there is a transition from local
to global oscillations for all values of $\chi$. For large $\chi$, the three
species global oscillations phase extends up to $Q=1$. The amount of
long range interactions necessary to synchronize the oscillations, in 
this phase, decrease with $\chi$: the more homogeneous the RSP food web is,
the easier it is to synchronize distant regions. On the contrary, for 
small $\chi$ and above the local-global transition, the four species phase 
has another transition, to a homogeneous single species state (and,
obviously, no oscillations). For $\chi=0$ the amount of long range 
connections necessary to present global oscillations is minimum
because the food web, in this case, is a single loop. As $\chi$ increases inside
this phase, one has $d\Qc/d\chi>0$ until a maximum at the four-three species
transition: the larger the level of transitivity of the food web is,
the harder it is to synchronize. In other words, at the edge of the 4-3 transition,
local oscillations are quite robust and a greater amount of long range connections
are necessary in order to correlate different regions of the system. Away from
this point, either as $\chi\to 0$ or $\chi\to 1$, that corresponds to the most
homogeneous conditions, the amount of long range connections is minimum. Another
interesting feature of the phase diagram is that for the three species phase
it is harder to get the global synchronization state as the necessary fraction
of long range interactions is more than three times higher than in the four
species phase. Also, for $0.067<Q<0.093$, the local-global transition line is 
reentrant as a function of $\chi$.

\section{Conclusions}
\label{section.conclusions}

Transitivity and predation asymmetry are topological properties of oriented 
food webs that may prevent long term, stable coexistence within a population.
Both imply that the species are not all equivalent, but while the former is
characterized by closed loops not being cyclic, the latter has species with a different
number of preys and predators.  Full intransitivity and symmetry ($\chi=0$) are, 
however, not always required as is exemplified by the four species coexistence in
the model introduced in Ref.~\cite{LuRiAr13}, and studied both in mean field and, 
through simulations, on a regular lattice. 
%In this case, in which only short range interactions are present, same
%species individuals group into protective clusters while all competition between
%different species occur in a layer at the interface between these clusters.
Locally, each site follows a periodic sequence of strategies allowed in the food web
but these oscillations do not synchronize over long distances. Here we have
shown that upon introducing a finite fraction $Q$ of long range interactions, well 
separated regions may synchronize their local cycles and generate global density oscillations 
through a Hopf bifurcation even if the food web presents, in addition
to multiple loops, some amount of transitivity and asymmetry. 
%For $Q>0$, predators that were previously 
%constrained to attack only those preys on the surface of clusters, are now granted 
%direct access to their bulk.
For all values of $\chi$ there is a transition from local to global oscillations occurring 
inside the coexistence phase. The behavior is different for small or large $\chi$. For
large $\chi$, species 2 always gets extinct and the remaining three species phase,
reminiscent of the $\chi=1$ case, is stable irrespective of the value of $Q$. Interestingly,
the amount of long range connections necessary to synchronize the whole system decreases
the larger $\chi$ is, that is, the more homogeneous the RSP cycle is.
For $\chi=0$, on the other hand, the only possible phases have either four or one species:
in the absence of crossed interactions, once one species gets extinct, the food web is
no longer cyclic and only the prey of the extinct species remains. This transition, from
the four species coexistence to the absorbing state extends also for small $\chi$
and occurs when the amplitude of the limit cycle approaches the heteroclinic orbit.
Interestingly enough, the amount of long range interactions necessary to obtain
global synchronization, $\Qc(\chi)$, increases with $\chi$ inside the four species
phase: the more transitive are some of the subloops, the harder it is to synchronize
distant regions. Again one has that the less homogeneous is the food web, the larger
is the amount of long range interactions necessary to obtain synchronization.
Thus, $\Qc(\chi)$ presents a maximum close to the transition from 
three to four species, and it is harder to globally synchronize the density oscillations. 
Moreover, when all the four species are present 
and the food web is more complex, the synchronization transition requires 
a smaller amount of long range interactions than for three species (and a simpler
food web), that is, $\Qc(\chi)$ is smaller in the four species phase. 

At the two extremes of the phase diagram, $\chi=0$ and 1, the system reduces
to a single loop food web with four and three species, respectively. The local to global
transition occurs at $\Qc(\chi=0)\simeq 0.018$ and $\Qc(\chi=1)\simeq 0.068$, that
is, the threshold for ${\cal S}=4$ is smaller than for ${\cal S}=3$. Whether $\Qc$ 
further decreases with increasing ${\cal S}$, approaching an effectively short range
system with global oscillations as $\Qc\to 0$, and the interplay with the fixation that may happen for large 
${\cal S}$ are interesting questions that deserve further investigation.

%When four species global oscillations exist, the system follows the outer loop of the
%food web, 0123. It would be interesting to measure the amount of diversity on a
%local level since several subloops are possible and depend on $\chi$.
A first step towards the understanding of the complex behavior of large, cyclic trophic 
rings is done through simplified models going beyond the standard RSP game, as the one considered here. 
%For these
%more complex graphs, several relevant issues should be systematically addressed. For example, how robust
%is the biodiversity when the range of interactions is global but restricted~\cite{ZhChQiQi09} or the large 
%population viscosity constraint is relaxed and individuals are able to move~\cite{ReMoFr07a,ZhChQiQi09}? 
%In addition, t
The very existence of the local to
global oscillations transition, under broad conditions, in three and four species model, 
hints to the possibility that this is a general feature even in more
complex, realistic food webs. Thus, it would be important to study systems with more species (${\cal S}>4$)
in order to confirm the universality claim of the local to global oscillations phase.

%quantos estados absorventes este modelo tem? a transição 4-3 está em qual
%classe de universalidade (ver Szabo-Sznaider 2004)? DP? 

\begin{acknowledgments}
We thank A. F. Lütz for discussions and for a critical reading of the
manuscript. JJA thanks A. Szolnoki and G. Szabó for a very fruitful exchange of emails,
L. Brunnet and A. Endler for discussions on Hopf bifurcations 
and the INCT Sistemas Complexos and the Brazilian agencies CNPq, Fapergs and 
CAPES for partial support. CR thanks the IF-UFRGS for the hospitality 
and the ENS-Paris and Île de France (AMIE grant) for partial 
support during her stay in Porto Alegre.
\end{acknowledgments}

\bibliographystyle{apsrev}

%\bibliography{rsp4}

\input{rsp4cha.bbl}

\end{document}

%% file: rsp4cha.bbl

%% file: rsp4cha.bbl
\begin{thebibliography}{53}%
\makeatletter
\providecommand \@ifxundefined [1]{%
 \@ifx{#1\undefined}
}%
\providecommand \@ifnum [1]{%
 \ifnum #1\expandafter \@firstoftwo
 \else \expandafter \@secondoftwo
 \fi
}%
\providecommand \@ifx [1]{%
 \ifx #1\expandafter \@firstoftwo
 \else \expandafter \@secondoftwo
 \fi
}%
\providecommand \natexlab [1]{#1}%
\providecommand \enquote  [1]{``#1''}%
\providecommand \bibnamefont  [1]{#1}%
\providecommand \bibfnamefont [1]{#1}%
\providecommand \citenamefont [1]{#1}%
\providecommand \href@noop [0]{\@secondoftwo}%
\providecommand \href [0]{\begingroup \@sanitize@url \@href}%
\providecommand \@href[1]{\@@startlink{#1}\@@href}%
\providecommand \@@href[1]{\endgroup#1\@@endlink}%
\providecommand \@sanitize@url [0]{\catcode `\\12\catcode `\$12\catcode
  `\&12\catcode `\#12\catcode `\^12\catcode `\_12\catcode `\%12\relax}%
\providecommand \@@startlink[1]{}%
\providecommand \@@endlink[0]{}%
\providecommand \url  [0]{\begingroup\@sanitize@url \@url }%
\providecommand \@url [1]{\endgroup\@href {#1}{\urlprefix }}%
\providecommand \urlprefix  [0]{URL }%
\providecommand \Eprint [0]{\href }%
\providecommand \doibase [0]{http://dx.doi.org/}%
\providecommand \selectlanguage [0]{\@gobble}%
\providecommand \bibinfo  [0]{\@secondoftwo}%
\providecommand \bibfield  [0]{\@secondoftwo}%
\providecommand \translation [1]{[#1]}%
\providecommand \BibitemOpen [0]{}%
\providecommand \bibitemStop [0]{}%
\providecommand \bibitemNoStop [0]{.\EOS\space}%
\providecommand \EOS [0]{\spacefactor3000\relax}%
\providecommand \BibitemShut  [1]{\csname bibitem#1\endcsname}%
\let\auto@bib@innerbib\@empty
%</preamble>
\bibitem [{\citenamefont {Hofbauer}\ and\ \citenamefont
  {Sigmund}(1998)}]{HoSi98}%
  \BibitemOpen
  \bibfield  {author} {\bibinfo {author} {\bibfnamefont {J.}~\bibnamefont
  {Hofbauer}}\ and\ \bibinfo {author} {\bibfnamefont {K.}~\bibnamefont
  {Sigmund}},\ }\href@noop {} {\emph {\bibinfo {title} {Evolutionary Games and
  Population Dynamics}}}\ (\bibinfo  {publisher} {Cambridge University Press},\
  \bibinfo {address} {Cambridge},\ \bibinfo {year} {1998})\BibitemShut
  {NoStop}%
\bibitem [{\citenamefont {Szabó}\ and\ \citenamefont {Fath}(2007)}]{SzFa07}%
  \BibitemOpen
  \bibfield  {author} {\bibinfo {author} {\bibfnamefont {G.}~\bibnamefont
  {Szabó}}\ and\ \bibinfo {author} {\bibfnamefont {G.}~\bibnamefont {Fath}},\
  }\href@noop {} {\bibfield  {journal} {\bibinfo  {journal} {Phys. Rep.}\
  }\textbf {\bibinfo {volume} {446}},\ \bibinfo {pages} {97} (\bibinfo {year}
  {2007})}\BibitemShut {NoStop}%
\bibitem [{\citenamefont {Gilpin}(1975)}]{Gilpin75}%
  \BibitemOpen
  \bibfield  {author} {\bibinfo {author} {\bibfnamefont {M.~E.}\ \bibnamefont
  {Gilpin}},\ }\href@noop {} {\bibfield  {journal} {\bibinfo  {journal} {Am.
  Nat.}\ }\textbf {\bibinfo {volume} {109}},\ \bibinfo {pages} {51} (\bibinfo
  {year} {1975})}\BibitemShut {NoStop}%
\bibitem [{\citenamefont {Tainaka}(1988)}]{Tainaka88}%
  \BibitemOpen
  \bibfield  {author} {\bibinfo {author} {\bibfnamefont {K.-I.}\ \bibnamefont
  {Tainaka}},\ }\href@noop {} {\bibfield  {journal} {\bibinfo  {journal} {J.
  Phys. Soc. Japan}\ }\textbf {\bibinfo {volume} {57}},\ \bibinfo {pages}
  {2588} (\bibinfo {year} {1988})}\BibitemShut {NoStop}%
\bibitem [{\citenamefont {Sinervo}\ and\ \citenamefont
  {Lively}(1996)}]{SiLi96}%
  \BibitemOpen
  \bibfield  {author} {\bibinfo {author} {\bibfnamefont {B.}~\bibnamefont
  {Sinervo}}\ and\ \bibinfo {author} {\bibfnamefont {C.}~\bibnamefont
  {Lively}},\ }\href@noop {} {\bibfield  {journal} {\bibinfo  {journal}
  {Nature}\ }\textbf {\bibinfo {volume} {380}},\ \bibinfo {pages} {240}
  (\bibinfo {year} {1996})}\BibitemShut {NoStop}%
\bibitem [{\citenamefont {Kerr}\ \emph {et~al.}(2002)\citenamefont {Kerr},
  \citenamefont {Riley}, \citenamefont {Feldman},\ and\ \citenamefont
  {Bohannan}}]{KeRiFeBo02}%
  \BibitemOpen
  \bibfield  {author} {\bibinfo {author} {\bibfnamefont {B.}~\bibnamefont
  {Kerr}}, \bibinfo {author} {\bibfnamefont {M.~A.}\ \bibnamefont {Riley}},
  \bibinfo {author} {\bibfnamefont {M.~W.}\ \bibnamefont {Feldman}}, \ and\
  \bibinfo {author} {\bibfnamefont {B.~J.~M.}\ \bibnamefont {Bohannan}},\
  }\href@noop {} {\bibfield  {journal} {\bibinfo  {journal} {Nature}\ }\textbf
  {\bibinfo {volume} {418}},\ \bibinfo {pages} {171} (\bibinfo {year}
  {2002})}\BibitemShut {NoStop}%
\bibitem [{\citenamefont {Kirkup}\ and\ \citenamefont {Riley}(2004)}]{KiRi04}%
  \BibitemOpen
  \bibfield  {author} {\bibinfo {author} {\bibfnamefont {B.~C.}\ \bibnamefont
  {Kirkup}}\ and\ \bibinfo {author} {\bibfnamefont {M.~A.}\ \bibnamefont
  {Riley}},\ }\href@noop {} {\bibfield  {journal} {\bibinfo  {journal}
  {Nature}\ }\textbf {\bibinfo {volume} {428}},\ \bibinfo {pages} {412}
  (\bibinfo {year} {2004})}\BibitemShut {NoStop}%
\bibitem [{\citenamefont {Hibbing}\ \emph {et~al.}(2010)\citenamefont
  {Hibbing}, \citenamefont {Fuqua}, \citenamefont {Parsek},\ and\ \citenamefont
  {Peterson}}]{HiFuPaPe10}%
  \BibitemOpen
  \bibfield  {author} {\bibinfo {author} {\bibfnamefont {M.~E.}\ \bibnamefont
  {Hibbing}}, \bibinfo {author} {\bibfnamefont {C.}~\bibnamefont {Fuqua}},
  \bibinfo {author} {\bibfnamefont {M.~R.}\ \bibnamefont {Parsek}}, \ and\
  \bibinfo {author} {\bibfnamefont {S.~B.}\ \bibnamefont {Peterson}},\
  }\href@noop {} {\bibfield  {journal} {\bibinfo  {journal} {Nature Reviews:
  Microbiology}\ }\textbf {\bibinfo {volume} {8}},\ \bibinfo {pages} {15}
  (\bibinfo {year} {2010})}\BibitemShut {NoStop}%
\bibitem [{\citenamefont {Trosvik}\ \emph {et~al.}(2010)\citenamefont
  {Trosvik}, \citenamefont {Rudi}, \citenamefont {Strætkvern}, \citenamefont
  {Jakobsen}, \citenamefont {Næs},\ and\ \citenamefont
  {Stenseth}}]{Trosvik10}%
  \BibitemOpen
  \bibfield  {author} {\bibinfo {author} {\bibfnamefont {P.}~\bibnamefont
  {Trosvik}}, \bibinfo {author} {\bibfnamefont {K.}~\bibnamefont {Rudi}},
  \bibinfo {author} {\bibfnamefont {K.~O.}\ \bibnamefont {Strætkvern}},
  \bibinfo {author} {\bibfnamefont {K.~S.}\ \bibnamefont {Jakobsen}}, \bibinfo
  {author} {\bibfnamefont {T.}~\bibnamefont {Næs}}, \ and\ \bibinfo {author}
  {\bibfnamefont {N.~C.}\ \bibnamefont {Stenseth}},\ }\href@noop {} {\bibfield
  {journal} {\bibinfo  {journal} {Environ. Microb.}\ }\textbf {\bibinfo
  {volume} {12}},\ \bibinfo {pages} {2677} (\bibinfo {year}
  {2010})}\BibitemShut {NoStop}%
\bibitem [{\citenamefont {Buss}\ and\ \citenamefont {Jackson}(1979)}]{BuJa79}%
  \BibitemOpen
  \bibfield  {author} {\bibinfo {author} {\bibfnamefont {L.~W.}\ \bibnamefont
  {Buss}}\ and\ \bibinfo {author} {\bibfnamefont {J.~B.~C.}\ \bibnamefont
  {Jackson}},\ }\href@noop {} {\bibfield  {journal} {\bibinfo  {journal} {Am.
  Nat.}\ }\textbf {\bibinfo {volume} {113}},\ \bibinfo {pages} {223} (\bibinfo
  {year} {1979})}\BibitemShut {NoStop}%
\bibitem [{\citenamefont {Watt}(1947)}]{Watt47}%
  \BibitemOpen
  \bibfield  {author} {\bibinfo {author} {\bibfnamefont {A.~S.}\ \bibnamefont
  {Watt}},\ }\href@noop {} {\bibfield  {journal} {\bibinfo  {journal} {J.
  Ecol.}\ }\textbf {\bibinfo {volume} {35}},\ \bibinfo {pages} {1} (\bibinfo
  {year} {1947})}\BibitemShut {NoStop}%
\bibitem [{\citenamefont {Th\'orhallsd\'ottir}(1990)}]{Thorhallsdottir90}%
  \BibitemOpen
  \bibfield  {author} {\bibinfo {author} {\bibfnamefont {T.~E.}\ \bibnamefont
  {Th\'orhallsd\'ottir}},\ }\href@noop {} {\bibfield  {journal} {\bibinfo
  {journal} {J. Ecol.}\ }\textbf {\bibinfo {volume} {78}},\ \bibinfo {pages}
  {909} (\bibinfo {year} {1990})}\BibitemShut {NoStop}%
\bibitem [{\citenamefont {Silvertown}\ \emph {et~al.}(1994)\citenamefont
  {Silvertown}, \citenamefont {Lines},\ and\ \citenamefont {Dale}}]{SiLiDa94}%
  \BibitemOpen
  \bibfield  {author} {\bibinfo {author} {\bibfnamefont {J.}~\bibnamefont
  {Silvertown}}, \bibinfo {author} {\bibfnamefont {C.~E.~M.}\ \bibnamefont
  {Lines}}, \ and\ \bibinfo {author} {\bibfnamefont {M.~P.}\ \bibnamefont
  {Dale}},\ }\href@noop {} {\bibfield  {journal} {\bibinfo  {journal} {J.
  Ecol.}\ }\textbf {\bibinfo {volume} {82}},\ \bibinfo {pages} {31} (\bibinfo
  {year} {1994})}\BibitemShut {NoStop}%
\bibitem [{\citenamefont {Frachebourg}\ \emph {et~al.}(1996)\citenamefont
  {Frachebourg}, \citenamefont {Krapivsky},\ and\ \citenamefont
  {Ben-Naim}}]{FrKrBe96}%
  \BibitemOpen
  \bibfield  {author} {\bibinfo {author} {\bibfnamefont {L.}~\bibnamefont
  {Frachebourg}}, \bibinfo {author} {\bibfnamefont {P.~L.}\ \bibnamefont
  {Krapivsky}}, \ and\ \bibinfo {author} {\bibfnamefont {E.}~\bibnamefont
  {Ben-Naim}},\ }\href@noop {} {\bibfield  {journal} {\bibinfo  {journal}
  {Phys. Rev. E}\ }\textbf {\bibinfo {volume} {54}},\ \bibinfo {pages} {6186}
  (\bibinfo {year} {1996})}\BibitemShut {NoStop}%
\bibitem [{\citenamefont {Frachebourg}\ and\ \citenamefont
  {Krapivsky}(1998)}]{FrKr98}%
  \BibitemOpen
  \bibfield  {author} {\bibinfo {author} {\bibfnamefont {L.}~\bibnamefont
  {Frachebourg}}\ and\ \bibinfo {author} {\bibfnamefont {P.~L.}\ \bibnamefont
  {Krapivsky}},\ }\href@noop {} {\bibfield  {journal} {\bibinfo  {journal} {J.
  Phys. A: Math. Gen.}\ }\textbf {\bibinfo {volume} {31}},\ \bibinfo {pages}
  {L287} (\bibinfo {year} {1998})}\BibitemShut {NoStop}%
\bibitem [{\citenamefont {Sato}\ \emph {et~al.}(2002)\citenamefont {Sato},
  \citenamefont {Yoshida},\ and\ \citenamefont {Konno}}]{SaYoKo02}%
  \BibitemOpen
  \bibfield  {author} {\bibinfo {author} {\bibfnamefont {K.}~\bibnamefont
  {Sato}}, \bibinfo {author} {\bibfnamefont {N.}~\bibnamefont {Yoshida}}, \
  and\ \bibinfo {author} {\bibfnamefont {N.}~\bibnamefont {Konno}},\
  }\href@noop {} {\bibfield  {journal} {\bibinfo  {journal} {Appl. Math.
  Comput.}\ }\textbf {\bibinfo {volume} {126}},\ \bibinfo {pages} {255}
  (\bibinfo {year} {2002})}\BibitemShut {NoStop}%
\bibitem [{\citenamefont {Szabó}\ and\ \citenamefont
  {Sznaider}(2004)}]{SzSz04b}%
  \BibitemOpen
  \bibfield  {author} {\bibinfo {author} {\bibfnamefont {G.}~\bibnamefont
  {Szabó}}\ and\ \bibinfo {author} {\bibfnamefont {G.~A.}\ \bibnamefont
  {Sznaider}},\ }\href@noop {} {\bibfield  {journal} {\bibinfo  {journal}
  {Phys. Rev. E}\ }\textbf {\bibinfo {volume} {69}},\ \bibinfo {pages} {031911}
  (\bibinfo {year} {2004})}\BibitemShut {NoStop}%
\bibitem [{\citenamefont {Szabó}\ \emph {et~al.}(2007)\citenamefont {Szabó},
  \citenamefont {Szolnoki},\ and\ \citenamefont {Sznaider}}]{SzSzSz07}%
  \BibitemOpen
  \bibfield  {author} {\bibinfo {author} {\bibfnamefont {G.}~\bibnamefont
  {Szabó}}, \bibinfo {author} {\bibfnamefont {A.}~\bibnamefont {Szolnoki}}, \
  and\ \bibinfo {author} {\bibfnamefont {G.~A.}\ \bibnamefont {Sznaider}},\
  }\href@noop {} {\bibfield  {journal} {\bibinfo  {journal} {Phys. Rev. E}\
  }\textbf {\bibinfo {volume} {76}},\ \bibinfo {pages} {051921} (\bibinfo
  {year} {2007})}\BibitemShut {NoStop}%
\bibitem [{\citenamefont {Case}\ \emph {et~al.}(2010)\citenamefont {Case},
  \citenamefont {Durney}, \citenamefont {Pleimling},\ and\ \citenamefont
  {R.K.P.Zia}}]{CaDuPlZi10}%
  \BibitemOpen
  \bibfield  {author} {\bibinfo {author} {\bibfnamefont {S.~O.}\ \bibnamefont
  {Case}}, \bibinfo {author} {\bibfnamefont {C.~H.}\ \bibnamefont {Durney}},
  \bibinfo {author} {\bibfnamefont {M.}~\bibnamefont {Pleimling}}, \ and\
  \bibinfo {author} {\bibnamefont {R.K.P.Zia}},\ }\href@noop {} {\bibfield
  {journal} {\bibinfo  {journal} {EPL}\ }\textbf {\bibinfo {volume} {92}},\
  \bibinfo {pages} {58003} (\bibinfo {year} {2010})}\BibitemShut {NoStop}%
\bibitem [{\citenamefont {Durney}\ \emph {et~al.}(2011)\citenamefont {Durney},
  \citenamefont {Case}, \citenamefont {Pleimling},\ and\ \citenamefont
  {P.Zia}}]{DuCaPlZi11}%
  \BibitemOpen
  \bibfield  {author} {\bibinfo {author} {\bibfnamefont {C.~H.}\ \bibnamefont
  {Durney}}, \bibinfo {author} {\bibfnamefont {S.~O.}\ \bibnamefont {Case}},
  \bibinfo {author} {\bibfnamefont {M.}~\bibnamefont {Pleimling}}, \ and\
  \bibinfo {author} {\bibfnamefont {R.~K.}\ \bibnamefont {P.Zia}},\ }\href@noop
  {} {\bibfield  {journal} {\bibinfo  {journal} {Phys. Rev. E}\ }\textbf
  {\bibinfo {volume} {83}},\ \bibinfo {pages} {051108} (\bibinfo {year}
  {2011})}\BibitemShut {NoStop}%
\bibitem [{\citenamefont {Roman}\ \emph {et~al.}(2012)\citenamefont {Roman},
  \citenamefont {Konrad},\ and\ \citenamefont {Pleimling}}]{RoKoPl12}%
  \BibitemOpen
  \bibfield  {author} {\bibinfo {author} {\bibfnamefont {A.}~\bibnamefont
  {Roman}}, \bibinfo {author} {\bibfnamefont {D.}~\bibnamefont {Konrad}}, \
  and\ \bibinfo {author} {\bibfnamefont {M.}~\bibnamefont {Pleimling}},\
  }\href@noop {} {\bibfield  {journal} {\bibinfo  {journal} {J. Stat. Mech.}\
  ,\ \bibinfo {pages} {P07014}} (\bibinfo {year} {2012})}\BibitemShut {NoStop}%
\bibitem [{\citenamefont {Szabó}(2005)}]{Szabo05}%
  \BibitemOpen
  \bibfield  {author} {\bibinfo {author} {\bibfnamefont {G.}~\bibnamefont
  {Szabó}},\ }\href@noop {} {\bibfield  {journal} {\bibinfo  {journal} {J.
  Phys. A: Math. Gen.}\ }\textbf {\bibinfo {volume} {38}},\ \bibinfo {pages}
  {6689} (\bibinfo {year} {2005})}\BibitemShut {NoStop}%
\bibitem [{\citenamefont {Szab\'o}\ and\ \citenamefont
  {Szolnoki}(2008)}]{SzSz08}%
  \BibitemOpen
  \bibfield  {author} {\bibinfo {author} {\bibfnamefont {G.}~\bibnamefont
  {Szab\'o}}\ and\ \bibinfo {author} {\bibfnamefont {A.}~\bibnamefont
  {Szolnoki}},\ }\href@noop {} {\bibfield  {journal} {\bibinfo  {journal}
  {Phys. Rev. E}\ }\textbf {\bibinfo {volume} {77}},\ \bibinfo {pages} {011906}
  (\bibinfo {year} {2008})}\BibitemShut {NoStop}%
\bibitem [{\citenamefont {Dobrinevski}\ and\ \citenamefont
  {Frey}(2012)}]{DoFr12}%
  \BibitemOpen
  \bibfield  {author} {\bibinfo {author} {\bibfnamefont {A.}~\bibnamefont
  {Dobrinevski}}\ and\ \bibinfo {author} {\bibfnamefont {E.}~\bibnamefont
  {Frey}},\ }\href@noop {} {\bibfield  {journal} {\bibinfo  {journal} {Phys.
  Rev. E}\ }\textbf {\bibinfo {volume} {85}},\ \bibinfo {pages} {051903}
  (\bibinfo {year} {2012})}\BibitemShut {NoStop}%
\bibitem [{\citenamefont {Guisoni}\ \emph {et~al.}(2013)\citenamefont
  {Guisoni}, \citenamefont {Loscar},\ and\ \citenamefont {Girardi}}]{GuLoGi13}%
  \BibitemOpen
  \bibfield  {author} {\bibinfo {author} {\bibfnamefont {N.~C.}\ \bibnamefont
  {Guisoni}}, \bibinfo {author} {\bibfnamefont {E.~S.}\ \bibnamefont {Loscar}},
  \ and\ \bibinfo {author} {\bibfnamefont {M.}~\bibnamefont {Girardi}},\
  }\href@noop {} {\bibfield  {journal} {\bibinfo  {journal} {Phys. Rev. E}\
  }\textbf {\bibinfo {volume} {88}},\ \bibinfo {pages} {022133} (\bibinfo
  {year} {2013})}\BibitemShut {NoStop}%
\bibitem [{\citenamefont {Intoy}\ and\ \citenamefont
  {Pleimling}(2013)}]{InPl13}%
  \BibitemOpen
  \bibfield  {author} {\bibinfo {author} {\bibfnamefont {B.}~\bibnamefont
  {Intoy}}\ and\ \bibinfo {author} {\bibfnamefont {M.}~\bibnamefont
  {Pleimling}},\ }\href@noop {} {\bibfield  {journal} {\bibinfo  {journal} {J.
  Stat. Mech.}\ }\textbf {\bibinfo {volume} {2013}},\ \bibinfo {pages} {P08011}
  (\bibinfo {year} {2013})}\BibitemShut {NoStop}%
\bibitem [{Note1()}]{Note1}%
  \BibitemOpen
  \bibinfo {note} {Differently from a passive defensive alliance, the species
  belonging to an active alliance also predate on each other, but prevent
  species not belonging to the subcycle from invading.}\BibitemShut {Stop}%
\bibitem [{\citenamefont {Silvertown}\ \emph {et~al.}(1992)\citenamefont
  {Silvertown}, \citenamefont {Holtier}, \citenamefont {Johnson},\ and\
  \citenamefont {Dale}}]{SiHoJoDa92}%
  \BibitemOpen
  \bibfield  {author} {\bibinfo {author} {\bibfnamefont {J.}~\bibnamefont
  {Silvertown}}, \bibinfo {author} {\bibfnamefont {S.}~\bibnamefont {Holtier}},
  \bibinfo {author} {\bibfnamefont {J.}~\bibnamefont {Johnson}}, \ and\
  \bibinfo {author} {\bibfnamefont {M.~P.}\ \bibnamefont {Dale}},\ }\href@noop
  {} {\bibfield  {journal} {\bibinfo  {journal} {J. Ecol.}\ }\textbf {\bibinfo
  {volume} {80}},\ \bibinfo {pages} {527} (\bibinfo {year} {1992})}\BibitemShut
  {NoStop}%
\bibitem [{\citenamefont {Durrett}\ and\ \citenamefont {Levin}(1998)}]{DuLe98}%
  \BibitemOpen
  \bibfield  {author} {\bibinfo {author} {\bibfnamefont {R.}~\bibnamefont
  {Durrett}}\ and\ \bibinfo {author} {\bibfnamefont {S.}~\bibnamefont
  {Levin}},\ }\href@noop {} {\bibfield  {journal} {\bibinfo  {journal} {Theor.
  Pop. Biol.}\ }\textbf {\bibinfo {volume} {53}},\ \bibinfo {pages} {30}
  (\bibinfo {year} {1998})}\BibitemShut {NoStop}%
\bibitem [{\citenamefont {Szabó}\ and\ \citenamefont
  {Cz\'ar\'an}(2001{\natexlab{a}})}]{SzCz01a}%
  \BibitemOpen
  \bibfield  {author} {\bibinfo {author} {\bibfnamefont {G.}~\bibnamefont
  {Szabó}}\ and\ \bibinfo {author} {\bibfnamefont {T.}~\bibnamefont
  {Cz\'ar\'an}},\ }\href@noop {} {\bibfield  {journal} {\bibinfo  {journal}
  {Phys. Rev. E}\ }\textbf {\bibinfo {volume} {64}},\ \bibinfo {pages} {042902}
  (\bibinfo {year} {2001}{\natexlab{a}})}\BibitemShut {NoStop}%
\bibitem [{\citenamefont {Szabó}\ and\ \citenamefont
  {Cz\'ar\'an}(2001{\natexlab{b}})}]{SzCz01b}%
  \BibitemOpen
  \bibfield  {author} {\bibinfo {author} {\bibfnamefont {G.}~\bibnamefont
  {Szabó}}\ and\ \bibinfo {author} {\bibfnamefont {T.}~\bibnamefont
  {Cz\'ar\'an}},\ }\href@noop {} {\bibfield  {journal} {\bibinfo  {journal}
  {Phys. Rev. E}\ }\textbf {\bibinfo {volume} {63}},\ \bibinfo {pages} {061904}
  (\bibinfo {year} {2001}{\natexlab{b}})}\BibitemShut {NoStop}%
\bibitem [{\citenamefont {Kim}\ \emph {et~al.}(2005)\citenamefont {Kim},
  \citenamefont {Liu}, \citenamefont {Um},\ and\ \citenamefont
  {Lee}}]{KiLiUmLe05}%
  \BibitemOpen
  \bibfield  {author} {\bibinfo {author} {\bibfnamefont {B.~J.}\ \bibnamefont
  {Kim}}, \bibinfo {author} {\bibfnamefont {J.}~\bibnamefont {Liu}}, \bibinfo
  {author} {\bibfnamefont {J.}~\bibnamefont {Um}}, \ and\ \bibinfo {author}
  {\bibfnamefont {S.-I.}\ \bibnamefont {Lee}},\ }\href@noop {} {\bibfield
  {journal} {\bibinfo  {journal} {Phys. Rev. E}\ }\textbf {\bibinfo {volume}
  {72}},\ \bibinfo {pages} {041906} (\bibinfo {year} {2005})}\BibitemShut
  {NoStop}%
\bibitem [{\citenamefont {Perc}\ \emph {et~al.}(2007)\citenamefont {Perc},
  \citenamefont {Szolnoki},\ and\ \citenamefont {Szab\'{o}}}]{PeSzSz07}%
  \BibitemOpen
  \bibfield  {author} {\bibinfo {author} {\bibfnamefont {M.}~\bibnamefont
  {Perc}}, \bibinfo {author} {\bibfnamefont {A.}~\bibnamefont {Szolnoki}}, \
  and\ \bibinfo {author} {\bibfnamefont {G.}~\bibnamefont {Szab\'{o}}},\
  }\href@noop {} {\bibfield  {journal} {\bibinfo  {journal} {Phys. Rev. E}\
  }\textbf {\bibinfo {volume} {75}},\ \bibinfo {eid} {052102} (\bibinfo {year}
  {2007})}\BibitemShut {NoStop}%
\bibitem [{\citenamefont {Szabó}\ \emph {et~al.}(2008)\citenamefont {Szabó},
  \citenamefont {Szolnoki},\ and\ \citenamefont {Borsos}}]{SzSzBo08}%
  \BibitemOpen
  \bibfield  {author} {\bibinfo {author} {\bibfnamefont {G.}~\bibnamefont
  {Szabó}}, \bibinfo {author} {\bibfnamefont {A.}~\bibnamefont {Szolnoki}}, \
  and\ \bibinfo {author} {\bibfnamefont {I.}~\bibnamefont {Borsos}},\
  }\href@noop {} {\bibfield  {journal} {\bibinfo  {journal} {Phys. Rev. E}\
  }\textbf {\bibinfo {volume} {77}},\ \bibinfo {pages} {041919} (\bibinfo
  {year} {2008})}\BibitemShut {NoStop}%
\bibitem [{\citenamefont {Laird}\ and\ \citenamefont {Schamp}(2008)}]{LaSc08}%
  \BibitemOpen
  \bibfield  {author} {\bibinfo {author} {\bibfnamefont {R.~A.}\ \bibnamefont
  {Laird}}\ and\ \bibinfo {author} {\bibfnamefont {B.~S.}\ \bibnamefont
  {Schamp}},\ }\href@noop {} {\bibfield  {journal} {\bibinfo  {journal}
  {Ecology}\ }\textbf {\bibinfo {volume} {89}},\ \bibinfo {pages} {237}
  (\bibinfo {year} {2008})}\BibitemShut {NoStop}%
\bibitem [{\citenamefont {Laird}\ and\ \citenamefont {Schamp}(2009)}]{LaSc09}%
  \BibitemOpen
  \bibfield  {author} {\bibinfo {author} {\bibfnamefont {R.~A.}\ \bibnamefont
  {Laird}}\ and\ \bibinfo {author} {\bibfnamefont {B.~S.}\ \bibnamefont
  {Schamp}},\ }\href@noop {} {\bibfield  {journal} {\bibinfo  {journal} {J.
  Theor. Biol.}\ }\textbf {\bibinfo {volume} {256}},\ \bibinfo {pages} {90}
  (\bibinfo {year} {2009})}\BibitemShut {NoStop}%
\bibitem [{\citenamefont {Han}\ \emph {et~al.}(2009)\citenamefont {Han},
  \citenamefont {Park},\ and\ \citenamefont {Kim}}]{HaPaKi09}%
  \BibitemOpen
  \bibfield  {author} {\bibinfo {author} {\bibfnamefont {S.-G.}\ \bibnamefont
  {Han}}, \bibinfo {author} {\bibfnamefont {S.-C.}\ \bibnamefont {Park}}, \
  and\ \bibinfo {author} {\bibfnamefont {B.~J.}\ \bibnamefont {Kim}},\
  }\href@noop {} {\bibfield  {journal} {\bibinfo  {journal} {Phys. Rev. E}\
  }\textbf {\bibinfo {volume} {79}},\ \bibinfo {pages} {066114} (\bibinfo
  {year} {2009})}\BibitemShut {NoStop}%
\bibitem [{\citenamefont {Li}\ \emph {et~al.}(2012)\citenamefont {Li},
  \citenamefont {Dong},\ and\ \citenamefont {Yang}}]{LiDoYa12}%
  \BibitemOpen
  \bibfield  {author} {\bibinfo {author} {\bibfnamefont {Y.}~\bibnamefont
  {Li}}, \bibinfo {author} {\bibfnamefont {L.}~\bibnamefont {Dong}}, \ and\
  \bibinfo {author} {\bibfnamefont {G.}~\bibnamefont {Yang}},\ }\href@noop {}
  {\bibfield  {journal} {\bibinfo  {journal} {Physica A}\ }\textbf {\bibinfo
  {volume} {391}},\ \bibinfo {pages} {125} (\bibinfo {year}
  {2012})}\BibitemShut {NoStop}%
\bibitem [{\citenamefont {Avelino}\ \emph
  {et~al.}(2012{\natexlab{a}})\citenamefont {Avelino}, \citenamefont {Bazeia},
  \citenamefont {Losano},\ and\ \citenamefont {Menezes}}]{AvBaLoMe12}%
  \BibitemOpen
  \bibfield  {author} {\bibinfo {author} {\bibfnamefont {P.}~\bibnamefont
  {Avelino}}, \bibinfo {author} {\bibfnamefont {D.}~\bibnamefont {Bazeia}},
  \bibinfo {author} {\bibfnamefont {L.}~\bibnamefont {Losano}}, \ and\ \bibinfo
  {author} {\bibfnamefont {J.}~\bibnamefont {Menezes}},\ }\href@noop {}
  {\bibfield  {journal} {\bibinfo  {journal} {Phys. Rev. E}\ }\textbf {\bibinfo
  {volume} {86}},\ \bibinfo {pages} {031119} (\bibinfo {year}
  {2012}{\natexlab{a}})}\BibitemShut {NoStop}%
\bibitem [{\citenamefont {Avelino}\ \emph
  {et~al.}(2012{\natexlab{b}})\citenamefont {Avelino}, \citenamefont {Bazeia},
  \citenamefont {Losano}, \citenamefont {Menezes},\ and\ \citenamefont
  {Oliveira}}]{AvBaLoMeOl12}%
  \BibitemOpen
  \bibfield  {author} {\bibinfo {author} {\bibfnamefont {P.}~\bibnamefont
  {Avelino}}, \bibinfo {author} {\bibfnamefont {D.}~\bibnamefont {Bazeia}},
  \bibinfo {author} {\bibfnamefont {L.}~\bibnamefont {Losano}}, \bibinfo
  {author} {\bibfnamefont {J.}~\bibnamefont {Menezes}}, \ and\ \bibinfo
  {author} {\bibfnamefont {B.}~\bibnamefont {Oliveira}},\ }\href@noop {}
  {\bibfield  {journal} {\bibinfo  {journal} {Phys. Rev. E}\ }\textbf {\bibinfo
  {volume} {86}},\ \bibinfo {pages} {036112} (\bibinfo {year}
  {2012}{\natexlab{b}})}\BibitemShut {NoStop}%
\bibitem [{\citenamefont {Roman}\ \emph {et~al.}(2013)\citenamefont {Roman},
  \citenamefont {Dasgupta},\ and\ \citenamefont {Pleimling}}]{RoDaPl13}%
  \BibitemOpen
  \bibfield  {author} {\bibinfo {author} {\bibfnamefont {A.}~\bibnamefont
  {Roman}}, \bibinfo {author} {\bibfnamefont {D.}~\bibnamefont {Dasgupta}}, \
  and\ \bibinfo {author} {\bibfnamefont {M.}~\bibnamefont {Pleimling}},\
  }\href@noop {} {\bibfield  {journal} {\bibinfo  {journal} {Phys. Rev. E}\
  }\textbf {\bibinfo {volume} {87}},\ \bibinfo {pages} {032148} (\bibinfo
  {year} {2013})}\BibitemShut {NoStop}%
\bibitem [{\citenamefont {L\"{u}tz}\ \emph {et~al.}(2013)\citenamefont
  {L\"{u}tz}, \citenamefont {Risau-Gusman},\ and\ \citenamefont
  {Arenzon}}]{LuRiAr13}%
  \BibitemOpen
  \bibfield  {author} {\bibinfo {author} {\bibfnamefont {A.~F.}\ \bibnamefont
  {L\"{u}tz}}, \bibinfo {author} {\bibfnamefont {S.}~\bibnamefont
  {Risau-Gusman}}, \ and\ \bibinfo {author} {\bibfnamefont {J.~J.}\
  \bibnamefont {Arenzon}},\ }\href@noop {} {\bibfield  {journal} {\bibinfo
  {journal} {J. Theor. Biol.}\ }\textbf {\bibinfo {volume} {317}},\ \bibinfo
  {pages} {286} (\bibinfo {year} {2013})}\BibitemShut {NoStop}%
\bibitem [{\citenamefont {Knebel}\ \emph {et~al.}(2013)\citenamefont {Knebel},
  \citenamefont {Kr\"{u}ger}, \citenamefont {Weber},\ and\ \citenamefont
  {Frey}}]{KnKrWeFr13}%
  \BibitemOpen
  \bibfield  {author} {\bibinfo {author} {\bibfnamefont {J.}~\bibnamefont
  {Knebel}}, \bibinfo {author} {\bibfnamefont {T.}~\bibnamefont {Kr\"{u}ger}},
  \bibinfo {author} {\bibfnamefont {M.~F.}\ \bibnamefont {Weber}}, \ and\
  \bibinfo {author} {\bibfnamefont {E.}~\bibnamefont {Frey}},\ }\href@noop {}
  {\bibfield  {journal} {\bibinfo  {journal} {Phys. Rev. Lett.}\ }\textbf
  {\bibinfo {volume} {110}},\ \bibinfo {pages} {168106} (\bibinfo {year}
  {2013})}\BibitemShut {NoStop}%
\bibitem [{\citenamefont {Kuperman}\ and\ \citenamefont
  {Abramson}(2001)}]{KuAb01}%
  \BibitemOpen
  \bibfield  {author} {\bibinfo {author} {\bibfnamefont {M.}~\bibnamefont
  {Kuperman}}\ and\ \bibinfo {author} {\bibfnamefont {G.}~\bibnamefont
  {Abramson}},\ }\href@noop {} {\bibfield  {journal} {\bibinfo  {journal}
  {Phys. Rev. Lett.}\ }\textbf {\bibinfo {volume} {86}},\ \bibinfo {pages}
  {2909} (\bibinfo {year} {2001})}\BibitemShut {NoStop}%
\bibitem [{\citenamefont {Szabó}\ \emph {et~al.}(2004)\citenamefont {Szabó},
  \citenamefont {Szolnoki},\ and\ \citenamefont {Izsák}}]{SzSzIz04}%
  \BibitemOpen
  \bibfield  {author} {\bibinfo {author} {\bibfnamefont {G.}~\bibnamefont
  {Szabó}}, \bibinfo {author} {\bibfnamefont {A.}~\bibnamefont {Szolnoki}}, \
  and\ \bibinfo {author} {\bibfnamefont {R.}~\bibnamefont {Izsák}},\
  }\href@noop {} {\bibfield  {journal} {\bibinfo  {journal} {J. Phys. A: Math.
  Gen.}\ }\textbf {\bibinfo {volume} {37}},\ \bibinfo {pages} {2599} (\bibinfo
  {year} {2004})}\BibitemShut {NoStop}%
\bibitem [{\citenamefont {Ying}\ \emph {et~al.}(2007)\citenamefont {Ying},
  \citenamefont {Hua},\ and\ \citenamefont {Wang}}]{YiHuWa07}%
  \BibitemOpen
  \bibfield  {author} {\bibinfo {author} {\bibfnamefont {C.-Y.}\ \bibnamefont
  {Ying}}, \bibinfo {author} {\bibfnamefont {D.-Y.}\ \bibnamefont {Hua}}, \
  and\ \bibinfo {author} {\bibfnamefont {L.-Y.}\ \bibnamefont {Wang}},\
  }\href@noop {} {\bibfield  {journal} {\bibinfo  {journal} {J. Phys. A: Math.
  Theor.}\ }\textbf {\bibinfo {volume} {40}},\ \bibinfo {pages} {4477}
  (\bibinfo {year} {2007})}\BibitemShut {NoStop}%
\bibitem [{\citenamefont {Rojas-Echenique}\ and\ \citenamefont
  {Allesina}(2011)}]{RoAl11}%
  \BibitemOpen
  \bibfield  {author} {\bibinfo {author} {\bibfnamefont {J.}~\bibnamefont
  {Rojas-Echenique}}\ and\ \bibinfo {author} {\bibfnamefont {S.}~\bibnamefont
  {Allesina}},\ }\href@noop {} {\bibfield  {journal} {\bibinfo  {journal}
  {Ecology}\ }\textbf {\bibinfo {volume} {92}},\ \bibinfo {pages} {1174}
  (\bibinfo {year} {2011})}\BibitemShut {NoStop}%
\bibitem [{\citenamefont {Vukov}\ \emph {et~al.}(2013)\citenamefont {Vukov},
  \citenamefont {Szolnoki},\ and\ \citenamefont {Szab\'{o}}}]{VuSzSz13}%
  \BibitemOpen
  \bibfield  {author} {\bibinfo {author} {\bibfnamefont {J.}~\bibnamefont
  {Vukov}}, \bibinfo {author} {\bibfnamefont {A.}~\bibnamefont {Szolnoki}}, \
  and\ \bibinfo {author} {\bibfnamefont {G.}~\bibnamefont {Szab\'{o}}},\
  }\href@noop {} {\bibfield  {journal} {\bibinfo  {journal} {Phys. Rev. E}\
  }\textbf {\bibinfo {volume} {88}},\ \bibinfo {pages} {022123} (\bibinfo
  {year} {2013})}\BibitemShut {NoStop}%
\bibitem [{\citenamefont {Antal}\ and\ \citenamefont
  {Scheuring}(2006)}]{AnSc06}%
  \BibitemOpen
  \bibfield  {author} {\bibinfo {author} {\bibfnamefont {T.}~\bibnamefont
  {Antal}}\ and\ \bibinfo {author} {\bibfnamefont {I.}~\bibnamefont
  {Scheuring}},\ }\href@noop {} {\bibfield  {journal} {\bibinfo  {journal}
  {Bull. Math. Biol.}\ }\textbf {\bibinfo {volume} {68}},\ \bibinfo {pages}
  {1923} (\bibinfo {year} {2006})}\BibitemShut {NoStop}%
\bibitem [{\citenamefont {Reichenbach}\ \emph {et~al.}(2007)\citenamefont
  {Reichenbach}, \citenamefont {Mobilia},\ and\ \citenamefont
  {Frey}}]{ReMoFr07a}%
  \BibitemOpen
  \bibfield  {author} {\bibinfo {author} {\bibfnamefont {T.}~\bibnamefont
  {Reichenbach}}, \bibinfo {author} {\bibfnamefont {M.}~\bibnamefont
  {Mobilia}}, \ and\ \bibinfo {author} {\bibfnamefont {E.}~\bibnamefont
  {Frey}},\ }\href@noop {} {\bibfield  {journal} {\bibinfo  {journal} {Nature}\
  }\textbf {\bibinfo {volume} {448}},\ \bibinfo {pages} {1046} (\bibinfo {year}
  {2007})}\BibitemShut {NoStop}%
\bibitem [{\citenamefont {Cremer}\ \emph {et~al.}(2009)\citenamefont {Cremer},
  \citenamefont {Reichenbach},\ and\ \citenamefont {Frey.}}]{CrReFr09b}%
  \BibitemOpen
  \bibfield  {author} {\bibinfo {author} {\bibfnamefont {J.}~\bibnamefont
  {Cremer}}, \bibinfo {author} {\bibfnamefont {T.}~\bibnamefont {Reichenbach}},
  \ and\ \bibinfo {author} {\bibfnamefont {E.}~\bibnamefont {Frey.}},\
  }\href@noop {} {\bibfield  {journal} {\bibinfo  {journal} {New J. Phys.}\
  }\textbf {\bibinfo {volume} {11}},\ \bibinfo {pages} {093029} (\bibinfo
  {year} {2009})}\BibitemShut {NoStop}%
\bibitem [{\citenamefont {Schütt}\ and\ \citenamefont
  {Claussen}(2010)}]{ScCl10b}%
  \BibitemOpen
  \bibfield  {author} {\bibinfo {author} {\bibfnamefont {M.}~\bibnamefont
  {Schütt}}\ and\ \bibinfo {author} {\bibfnamefont {J.~C.}\ \bibnamefont
  {Claussen}},\ }\href@noop {} {\enquote {\bibinfo {title} {Stabilization of
  biodiversity in the coevolutionary rock-paper-scissors game on complex
  networks},}\ } (\bibinfo {year} {2010}),\ \bibinfo {note}
  {arXiv:1003.2922}\BibitemShut {NoStop}%
\bibitem [{\citenamefont {Szolnoki}\ and\ \citenamefont
  {Szabó}(2004)}]{SzSz04a}%
  \BibitemOpen
  \bibfield  {author} {\bibinfo {author} {\bibfnamefont {A.}~\bibnamefont
  {Szolnoki}}\ and\ \bibinfo {author} {\bibfnamefont {G.}~\bibnamefont
  {Szabó}},\ }\href@noop {} {\bibfield  {journal} {\bibinfo  {journal} {Phys.
  Rev. E}\ }\textbf {\bibinfo {volume} {70}},\ \bibinfo {pages} {037102}
  (\bibinfo {year} {2004})}\BibitemShut {NoStop}%
\end{thebibliography}
